\documentclass[aps,twocolumn,showpacs,nofootinbib]{revtex4}
\usepackage{graphicx}
\usepackage{bm}

\def\be{\begin{equation}}
\def\ee{\end{equation}}
\def\bea{\begin{eqnarray}}
\def\eea{\end{eqnarray}}
\def\bear{\begin{array}}
\def\ear{\end{array}}
\def\raw{\rightarrow}

\begin{document}
\title{Neutral current coherent pion production}
\author{L. Alvarez-Ruso}
\author{L. S. Geng}
\author{M. J. \surname{Vicente Vacas}}
\affiliation{Departamento de F\'{\i}sica Te\'orica and IFIC,
Universidad de Valencia - CSIC;
Institutos de Investigaci\'on de Paterna, Aptdo. 22085, 46071 Valencia, Spain.}


\begin{abstract}

We investigate the neutrino induced coherent pion production reaction 
 at low and intermediate energies. The model 
includes pion, nucleon and $\Delta(1232)$ as the relevant hadronic 
degrees of freedom. Nuclear medium effects on the production mechanisms 
and pion distortion are taken into account. We obtain that the dominance 
of the $\Delta$ excitation holds due to large cancellations among the 
background contributions. We consider two sets of vector and 
axial-vector $N$-$\Delta$ transition form-factors, evidencing the strong 
sensitivity of the results to the axial coupling $C_5^A(0)$. 
The differences between neutrino and antineutrino cross sections, 
emerging from interference terms, are also discussed.

\end{abstract}

\pacs{25.30.Pt, 13.15.+g, 23.40.Bw}

\maketitle

\section{Introduction}

The study of weak coherent pion production in charged current (CC) and neutral current (NC) processes, is in the agenda of several current and 
future experiments~\cite{Hasegawa:2005td,Mahn:2006ac,Drakoulakos:2004gn,Ayres:2004js}. 
For instance, the SciBooNE 
detector~\cite{Mahn:2006ac} should be able to identify $\pi^0$'s emitted in the forward direction, where most of the 
coherent fraction of the total NC$\pi^0$ production is concentrated. MINER$\nu$A~\cite{Drakoulakos:2004gn} 
will collect data with high statistics, allowing for a clear separation between coherent and incoherent 
processes, and the comparison between neutrino and antineutrino cross sections. The measurements will be 
performed on C, Fe and Pb targets, so that the A dependence can be established. 
The NC reaction $\nu + A \raw \nu  + \pi^0 + A$ is of particular importance as a source of background in $\nu_e$ 
appearance oscillation  experiments~\cite{Aguilar-Arevalo:2007it,Ayres:2004js}.
The presently available data on coherent  NC$\pi^0$ production are scarce. They were obtained at 
$\langle E_\nu \rangle=2$~GeV~\cite{Faissner:1983ng} or 
higher~\cite{Isiksal:1984vh}, often with large error bars that complicate the discrimination between theoretical predictions.  

Many theoretical calculations used PCAC extrapolated to $q^2 \neq 0$~\cite{Rein:1982pf,Belkov:1986hn,Paschos:2005km}. Other studies focused on the neutrino-energy region around 1~GeV, where the excitation of the $\Delta(1232)$ resonance 
gives the dominant contribution to weak pion production on the nucleon. They have stressed the relevance of the modification of 
the $\Delta$ spectral function inside the nuclear medium~\cite{Kim:1996az,Kelkar:1996iv,Singh:2006bm,Alvarez-Ruso:2007tt}. 
Another important ingredient is the pion distortion. In Refs.~\cite{Rein:1982pf, Paschos:2005km}, it is  taken into account by factorizing the pion-nucleus elastic cross section. In a more general fashion, it can be incorporated in the amplitude by means of the 
distorted wave Born approximation, using a pion wave function obtained in the eikonal limit~\cite{Singh:2006bm} 
or by solving the Klein-Gordon equation with a realistic optical potential~\cite{Kelkar:1996iv,Alvarez-Ruso:2007tt}.          
            
In this article we extend our model of Ref.~\cite{Alvarez-Ruso:2007tt} by adding low energy background terms to  
$\Delta$ excitation, and apply it to the NC case. Background terms for the weak pion production on the nucleon were considered in the past in Refs.~\cite{Fogli:1979cz}, and more recently in Refs.~\cite{Sato:2003rq,Hernandez:2007qq}. In Ref.~\cite{Sato:2003rq} it was shown the importance of the background terms and 
the dressing of the $\Delta$ vertices due to the rescattering. Later, this model has been applied to incoherent pion production in nuclei~\cite{Szczerbinska:2006wk}. In Ref.~\cite{Hernandez:2007qq}
the background terms were derived using 
the non-linear $\sigma$ model for pions and nucleons. The work of Ref.~\cite{Hernandez:2007qq} and some recent 
lattice~\cite{Alexandrou:2006mc} and quark model~\cite{BarquillaCano:2007yk} calculations suggest smaller values for the 
axial-vector $N$-$\Delta$ transition form factors (FF) than those traditionally found in the 
literature\footnote{Note, however, that a direct comparison of theoretical with empirical FF is not 
straightforward due to the dressing of the $\Delta$ vertices as obtained in Ref.~\cite{Sato:2003rq}.}.
Therefore, we explore the 
sensitivity of the weak coherent pion production cross sections to different choices of FF in the energy range of 
MiniBooNE and K2K experiments.

\section{Theoretical model}
\label{sec:th}

\begin{figure}[b]
	\includegraphics[width=0.71\columnwidth]{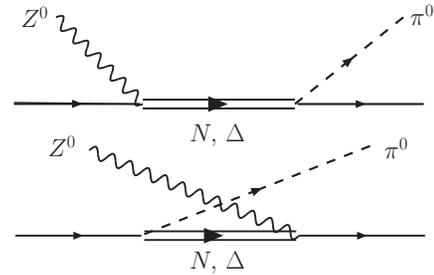}
	\caption{Mechanisms contributing to coherent $\pi^0$ production in isospin symmetric matter.}
	\label{fig:fig1}
\end{figure}

We have applied the model of Ref.~\cite{Hernandez:2007qq} for weak pion production on the nucleon to the description 
of the NC coherent reaction $\nu(k) + A \raw \nu(k')  + \pi^0(p_\pi) + A$
on (nearly) isospin symmetric nuclei such as $^{12}$C, $^{27}$Al and $^{56}$Fe.  
Besides the dominant direct $\Delta$ excitation~\cite{Leitner:2006sp}, this model includes the crossed $\Delta$ term and nucleon pole terms both direct and crossed shown    
in Fig.~\ref{fig:fig1}. 
The initial and final nucleons are in the same quantum state as required by coherence. 
Other possible contributions (see Fig. 2 of Ref.~\cite{Hernandez:2007qq} for the complete set of diagrams) 
cancel for isospin symmetric nuclei and will not be considered. 
The cross section for this reaction is given by 
\begin{equation}
\frac{d \sigma}{d \Omega_\nu d E_\nu d \Omega_\pi} = \frac{1}{8}
\frac{|\vec{k}'| |\vec{p}_\pi|} {|\vec{k}|} \frac{1}{(2 \pi)^5} \,
{{G^2}\over{2}}| l_{\mu} J^{\mu}_{A} |^2\,,
\end{equation}
where  the neutrino current 
$
l_{\mu}= \bar{u}_\nu (k') \gamma_{\mu} (1 -\gamma_{5}) u_{\nu}(k)\,.
$
The nuclear current $J^{\mu}_{A}$, obtained as the coherent sum over all nucleons, for all four mechanisms is
\bea
J^{\mu}_{\Delta d}&=&\frac{i}{\sqrt{6}} \frac{f^*}{m_\pi} \int d^3r \, e^{i\vec{q}\cdot \vec{r}}
\rho(r) F_\Delta(p_d) D_\Delta(p_d) \nonumber \\
&\times& 
\hat{p}_\pi^\alpha \mathrm{Tr}\left[ \bar{u}(0) \Lambda_{\alpha \beta}{\mathcal A}^{\beta \mu} u(0) \right] 
\phi^*_{>}(\vec{p}_\pi,\vec{r}) \,, \label{Fst}
\\ [0.2cm]
J^{\mu}_{\Delta c}&=& \frac{i}{\sqrt{6}} \frac{f^*}{m_\pi} \int d^3r \, e^{i\vec{q}\cdot \vec{r}}
\rho(r) F_\Delta(p_c) D_\Delta(p_c) \nonumber \\
&\times& 
\hat{p}_\pi^\alpha \mathrm{Tr}\left[ \bar{u}(0) \tilde{{\mathcal A}}^{\beta \mu} \Lambda_{\beta \alpha} u(0) \right] 
\phi^*_{>}(\vec{p}_\pi,\vec{r}) \,, 
\\ [0.2cm]
J^{\mu}_{N d}&=& -i \frac{g_A}{8 f_\pi} \int d^3r \, e^{i\vec{q}\cdot \vec{r}}
\rho(r) F_N(p_d) D_N(p_d) \nonumber \\
&\times& 
\hat{p}_\pi^\alpha \mathrm{Tr}\left[ \bar{u}(0) \gamma_\alpha \gamma_5 (p_d\!\!\!\!\!/+m_N) \mathcal{B}_V^\mu u(0) \right] 
\phi^*_{>}(\vec{p}_\pi,\vec{r}), 
\\ [0.2cm]
J^{\mu}_{N c}&=& -i \frac{g_A}{8 f_\pi} \int d^3r \, e^{i\vec{q}\cdot \vec{r}}
\rho(r) F_N(p_c) D_N(p_c) \nonumber \\
&\times& 
\hat{p}_\pi^\alpha \mathrm{Tr}\left[ \bar{u}(0) \mathcal{B}_V^\mu (p_d\!\!\!\!\!/+m_N) \gamma_\alpha \gamma_5  u(0) \right] 
\phi^*_{>}(\vec{p}_\pi,\vec{r}). \label{Lst}
\eea 
Here, $g_A=1.26$ is the axial coupling of the nucleon, $f_\pi=92.4$~MeV, the pion decay constant and 
$f^*=2.13$, the $\Delta \raw N \, \pi$ decay coupling; $m_\pi$ and $m_N$ stand for the pion and nucleon masses. 
The local nuclear density is denoted by $\rho (r)$. $D_\Delta(p') \Lambda_{\alpha \beta}(p')$ and $D_N(p') (p'\!\!\!\!/+m_N)$ are 
the $\Delta$ and nucleon propagators with the momentum of the intermediate state $p'$ being $p_d=p_i+q=p_f+p_\pi$ for the 
direct diagrams, and $p_c=p_i-p_\pi=p_f-q$ for the crossed ones; $p_i$ and $p_f$ are the initial and final nucleon momenta while $q$ is the 
momentum transfered by the lepton $q=k-k'$.  
The effect of hadron internal structure on the $\Delta N\pi$ and $NN\pi$ vertices has been parametrized as~\cite{Penner:2002ma} 
\be
F_{\Delta(N)}(p')=\frac{\Lambda^4}{\Lambda^4+(p'^2-m^2_{\Delta(N)})^2},
\ee 
with $\Lambda = 1$~GeV. In the evaluation of the spin traces, we neglect corrections of the order ($\vec{p}_{i(f)}/m_N$) which is equivalent 
to setting nucleon three-momenta to zero. On the other side, in the evaluation of $D_{\Delta(N)}(p')$, 
we assume that the three-momentum transfered to 
the nucleus is equally shared by the initial and final nucleons, so that 
$\vec{p}_{i(f)}=\pm (\vec{p}_\pi-\vec{q})/2$~\cite{Carrasco:1991we, Drechsel:1999vh}.     

The function $\mathcal{B}_V^\mu$ is given in terms of the vector and axial-vector nucleon form factors as follows
\be
\mathcal{B}_V^\mu = \tilde{F}_1^V \gamma^\mu + i \tilde{F}_2^V \sigma^{\mu \nu} \frac{q_\nu}{2 m_N} - \tilde{F}_A^V \gamma^\mu \gamma_5 - 
\tilde{F}_P^V q^\mu \gamma_5 \,,
\ee
where 
$
\tilde{F}_{1,2}^V =  (1-2 \sin^2{\theta_W}) (F_{1,2}^p - F_{1,2}^n)\,, \,\,
\tilde{F}_{A,P}^V = F_{A,P}
$.
We use the BBA-2003 parametrization~\cite{Budd:2003wb} for the vector FF, and the standard dipole 
$F_A = g_A (1 - q^2/M^2_A)^{-2}$
with $M_A=1$~GeV. PCAC has been applied to relate $F_P$ to $F_A$. 
${\mathcal A}^{\beta \mu}$ carries the information about the weak $N$-$\Delta$ transition 
($\tilde{{\mathcal A}}^{\beta \mu}(p',q)=\gamma_0 [{\mathcal A}^{\beta \mu}(p',-q)]^{\dagger} \gamma_0$) 
\bea
{\mathcal A}^{\beta\mu}= \left\{ 
  {{\tilde{C}_3^V}\over{m_N}} (g^{\beta \mu} q \!\!\! / - q^{\beta} \gamma^{\mu})+
  {{\tilde{C}_4^V}\over{m_N^2}} (g^{\beta \mu} q\cdot p' - q^{\beta} p'^{\mu}) \right. \nonumber \\  
  + \left. {{\tilde{C}_5^V}\over{m_N^2}} (g^{\beta \mu} q\cdot p - q^{\beta} p^{\mu})
   \right\}\gamma_{5}
  +\left\{ {{\tilde{C}_3^A}\over{m_N}} (g^{\beta \mu}q \!\!\! / - q^{\beta} \gamma^{\mu}) \right.
\nonumber \\
+ \left. {{\tilde{C}_4^A}\over{m_N^2}} (g^{\beta \mu} q\cdot p' - q^{\beta} p'^{\mu})+
 {\tilde{C}_5^A} g^{\beta \mu}+ {{\tilde{C}_6^A}\over{m_N^2}} q^{\beta} q^{\mu} \right\},  
\eea
where 
$
\tilde{C}_i^V = (1-2 \sin^2{\theta_W}) C_i^V $, 
$ 
\tilde{C}_i^A = C_i^A 
$.

For the FF $C_i^{V,A}$ we consider two different parametrizations. The first one, adopted in our recent study of CC coherent pion 
production~\cite{Alvarez-Ruso:2007tt},  assumes the $M_{1+}$ dominance of the 
$N$-$\Delta$ electromagnetic transition~\cite{Schreiner:1973mj}, and the Adler model~\cite{Adler:1968tw} for the axial FF, 
$C_4^A = - C_5^A /4 $, $ C_3^A=0$,
with 
\be
\label{setI}
C_5^A = {{C_5^A(0)\left( 1+{{1.21\, q^2}\over{2\,\,\mathrm{GeV}^2-q^2}} \right) }
{\left( 1- {{q^2}\over{M_{A\Delta}^2}}\right)^{-2}}}. 
\ee
We take $C_5^A(0) = 1.2$, in agreement with the off-diagonal Goldberger-Treiman relation~\cite{Schreiner:1973mj}, and 
$M_{A\Delta}=1.28$~GeV as extracted from BNL data~\cite{Kitagaki:1990vs}.  $C_6^A$ is related to $C_5^A$ by PCAC. 

The second set of FF is taken from Ref.~\cite{Hernandez:2007qq}. In the vector sector, it adopts the 
parametrizations of Ref.~\cite{Lalakulich:2005cs} obtained from a new analysis of electron-scattering data, which 
allows to go beyond the $M_{1+}$ approximation. In the axial sector, the Adler model is also used, but 
with a different ansatz for $C_5^A$~\cite{Paschos:2003qr}
\be
\label{setII}
C_5^A = C_5^A(0) \left(1 - \frac{q^2}{3 M_{A\Delta}^2}\right)^{-1} \left( 1- {{q^2}\over{M_{A\Delta}^2}}\right)^{-2} .
\ee   
Hernandez {\it et al.}~\cite{Hernandez:2007qq} fit the free parameters above to the ANL data~\cite{Radecky:1981fn} 
with a $\pi N$ invariant mass constraint of $W < 1.4$~GeV finding $M_{A\Delta}=0.985 \pm 0.082$~GeV and 
$C_5^A(0) = 0.867 \pm 0.075$. This value of  $C_5^A(0)$ is considerably smaller than the one obtained from the 
 off-diagonal Goldberger-Treiman relation. The coherent pion production process is more forward-peaked than 
the incoherent one, and filters part of the background terms. For these reasons, it is specially sensitive to 
$C_5^A(0)$ and, therefore, better suited to constrain its value than the incoherent reaction, provided that the 
experimental separation of  these two contributions is possible. 

The modification of the $\Delta$ properties inside the nuclei \cite{Singh:1998ha} causes a large reduction of the coherent pion production cross section, as 
shown in Refs.~\cite{Singh:2006bm,Alvarez-Ruso:2007tt}. Therefore, we take it into account for the direct $\Delta$ mechanism by adding 
a density dependent $\Delta$ selfenergy and modifying the free width in $D_\Delta(p_d)$, as explained in Section II~B of 
Ref.~\cite{Alvarez-Ruso:2007tt}. One would also expect some in-medium corrections to the amplitudes of the other diagrams but, as shown below, these mechanisms have only a very small impact on the observables. Thus, we neglect further modifications which 
would have a negligible effect.

Finally, the operator $\hat{p}^\alpha_\pi$ in Eqs.~(\ref{Fst}-\ref{Lst}) acts on  $\phi^*_{>}$ as 
\be
\hat{p}_\pi \phi^*_{>}(\vec{p}_\pi,\vec{r}) = (\omega_\pi \phi^*_{>}(\vec{p}_\pi,\vec{r}), i \vec{\nabla} \phi^*_{>}(\vec{p}_\pi,\vec{r}))\,.
\ee
 $\phi^*_{>}(\vec{p}_\pi,\vec{r})$ is the distorted wave function of the outgoing pion with asymptotic momentum $\vec{p}_\pi$, obtained as the solution of 
the Klein-Gordon equation with a microscopical optical potential based on the $\Delta$-hole model. Details can be found in 
Section II~D of Ref.~\cite{Alvarez-Ruso:2007tt} and in Refs.~\cite{Garcia-Recio:1989xa,Nieves:1991ye}.

\section{Results}
\label{sec:re}
\begin{figure}[hb]
	\includegraphics[height=\columnwidth,angle=-90]{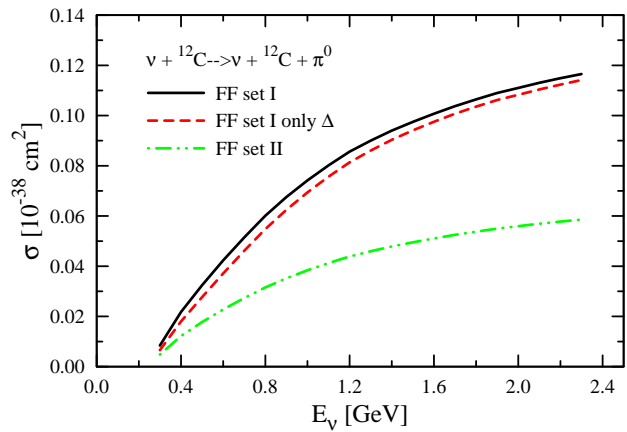}
	\caption{(Color online) Cross section for coherent $\pi^0$ production on $^{12}$C as a function of the neutrino energy.}
	\label{fig:fig2}
\end{figure}

The integrated cross section as a function of the neutrino energy is shown in Fig.~\ref{fig:fig2}. The comparison between the solid and the dashed 
lines, both obtained with the choice of FF denoted as set I, 
indicates that the direct $\Delta$ excitation is the dominant mechanism. The crossed $\Delta$ term 
accounts for about 0.6\% of the cross section at $E_\nu = 1$~GeV. The direct and crossed nucleon-pole terms are approximately 
six times bigger than the crossed $\Delta$, but there is a large cancellation between them, so that their sum gives less than 0.2\% of the total cross section. 
The more realistic description of the vector form factors of set II has 
a negligible effect on the result. On the contrary, the reduction of  $C_5^A(0)$ brings down the cross section by about a factor two with respect to set I. 
These features can be easily understood from the fact that in the forward direction ($q^2 = 0$), where most of the strength of this reaction is 
concentrated, the only form factor that contributes is $ C_5^A$~\cite{Alvarez-Ruso:1998hi}. Therefore, one can infer that
$
\sigma (\mathrm{I})/\sigma (\mathrm{II}) \sim \left[ C_{5(\mathrm{I})}^A(0)/C_{5(\mathrm{II})}^A(0)\right]^2 \approx 1.9 \,.
$    

\begin{figure}[ht]
	\includegraphics[height=\columnwidth,angle=-90]{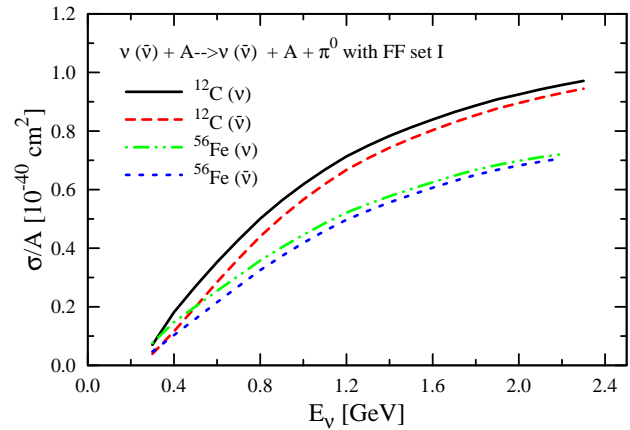}
	\caption{(Color online) $\nu$ \textit{vs.} $\bar{\nu}$ cross section for NC coherent $\pi^0$ production on $^{12}$C  and $^{56}$Fe as a function of  energy.}
	\label{fig:fig3}
\end{figure}

On the basis of this argument one also expects that interference plays a minor role. It is small indeed, but not negligible, as can 
be observed in Fig.~\ref{fig:fig3}. 
The formalism described in the previous section is applicable to antineutrinos replacing $(1 - \gamma_5)$ by $(1 + \gamma_5)$ in the 
leptonic current. 
The contributions of interference terms with $q^2 \neq 0$ are apparent in these curves, which would be otherwise identical for a given nucleus. 

A $\sigma(\bar{\nu}) \lesssim \sigma(\nu)$ is compatible with the findings of the Aachen-Padova collaboration on $^{27}$Al~\cite{Faissner:1983ng}. 
Our results, averaged over the CERN PS (anti)neutrino fluxes~\cite{Faissner:1977ku}, for both sets I and II are shown in Table~\ref{tab:table1} 
together with the experimental ones~\cite{Faissner:1983ng}. The results 
with set I are below the central experimental values but within the large error bars, while with set II the data are clearly underestimated. 
In any case, these theoretical values should be taken with care because these fluxes extend to high neutrino energies, where our model is less 
reliable.

We also show in Table \ref{tab:table1} our neutrino CC cross section on $^{12}$C averaged over the K2K flux~\cite{Ahn:2006zz}, and the (anti) neutrino NC ones 
averaged over the corresponding MiniBooNE fluxes~\cite{MiniBooNE}. 
In the case of K2K, the experimental threshold for the muon momentum 
$p_\mu >  450$~MeV/c is taken into account. The result with set I is slightly larger than the one given in our recent 
publication~\cite{Alvarez-Ruso:2007tt} ($10 \times 10^{-40}$~cm$^{2}$) due to the inclusion of background mechanisms. 
This number is above the upper bound obtained by K2K. While possible reasons 
for that discrepancy were discussed in Ref.~\cite{Alvarez-Ruso:2007tt}, here we find that the smaller $C_5^A(0)$ value obtained in 
Ref.~\cite{Hernandez:2007qq} (set II) is compatible with that limit. The situation is not so clear for MiniBooNE because set II predicts a very small
cross section although still within the large error bars of the preliminary experimental result. In the $\bar{\nu}$ mode, 
the value is smaller than in the $\nu$ mode, due to both the smaller antineutrino cross sections 
that we obtain  and the different energy distribution of the fluxes.     
\begin{table}
\caption{\label{tab:table1}Cross sections for weak coherent pion production in units of $10^{-40}\textrm{cm}^2$, 
averaged over the Aachen-Padova~(AP) \cite{Faissner:1977ku}, K2K~\cite{Ahn:2006zz} and MiniBooNE~\cite{MiniBooNE} spectra. 
Columns $\sigma_\mathrm{I(II)}$ correspond to the form factor sets I and II respectively.}
\begin{ruledtabular}
\begin{tabular}{lccccc}
Reaction & Experiment & $\sigma_\mathrm{I}$ & $\sigma_\mathrm{II}$  & $\sigma$ Exp.\\
\hline
NC $\nu +\, ^{27}$Al      & AP   &  19.9 & 10.1 &  29$\pm$ 10 \cite{Faissner:1983ng}\\
NC $\bar{\nu}+\, ^{27}$Al &  AP  & 19.7 & 9.8 & 25$\pm$ 7 \cite{Faissner:1983ng}\\
CC $\nu+\, ^{12}$C        & K2K & 10.8 & 5.7 & $< 7.7$ \cite{Hasegawa:2005td}\footnote{Obtained
using the ratio between coherent and $\sigma^{CC}$, the total CC cross section, and the value for $\sigma^{CC}$ of the K2K MC simulation.}\\
NC $\nu+\, ^{12}$C        & MiniBooNE  &  5.0   & 2.6 & $7.7 \pm 1.6\pm 3.6$\cite{Raaf:2005up}\footnote{Preliminary.}\\
NC $\bar{\nu}+\, ^{12}$C & MiniBooNE  &  4.6 & 2.2  &-&  
\end{tabular}
\end{ruledtabular}
\end{table} 

In summary,
we have calculated the NC coherent pion production cross sections at low energies and compared them with the still scarce data. Our model includes all mechanisms relevant for pion production at these energies. We find that the process is dominated by the excitation of the $\Delta$ resonance while the background terms play only a minor role. 
As in the CC case,
nuclear effects, mainly the $\Delta$ selfenergy and the pion distortion, produce a large reduction of the cross section.
The largest uncertainty of our results comes from the $N$-$\Delta$ FF. In order to estimate it we have presented results for two representative sets.
Actually, we find that
the coherent $\pi$ cross section is more sensitive to the $N$-$\Delta$ axial FF than the incoherent one. This is  due to the cancellation  of the isovector terms in isospin symmetric nuclei and  to the forward peaked kinematics driven by the nuclear form factor. We obtain smaller NC coherent $\pi$ production cross sections for neutrinos than for antineutrinos. Interference terms in the production amplitudes are responsible for this difference.

\begin{acknowledgments}
We thank M. Sorel and G. Zeller for useful discussions and valuable communications. 
This work was partially supported by the  MEC 
contract  FIS2006-03438, the Generalitat Valenciana ACOMP07/302,
and the EU Integrated Infrastructure
Initiative Hadron Physics Project contract RII3-CT-2004-506078.
\end{acknowledgments}

\bibliography{bibliocoh}

\end{document}